\documentclass{article}
\usepackage{arxiv}

\usepackage{subfigure}

\usepackage[utf8]{inputenc}
\usepackage{algorithm} 
\usepackage{algpseudocode}

\usepackage{tikz}
\usetikzlibrary{backgrounds,fit,decorations.pathreplacing, shapes}  
\usepackage{tkz-graph}

\usepackage[super]{nth}
\usepackage{adjustbox}
\usepackage{nameref}
\usepackage{hyperref}

\tikzstyle{vertex}=[circle,draw=black, fill=white,sloped,minimum size=17pt,inner sep=5pt]

\usepackage{tikz}
\usetikzlibrary{decorations.pathreplacing,calc}

\usepackage{enumerate}
\usepackage{amsmath}
\usepackage{graphicx}
\usepackage{wrapfig}
\usepackage{lscape}
\usepackage{rotating}
\usepackage{epstopdf}

\usepackage{caption}
\usepackage{url}
\usepackage{hyperref}
\usepackage{float}
\usepackage[super]{nth}
\usepackage[labelfont=bf]{caption}
\usepackage[labelsep=period]{caption}
\usepackage{csquotes}
\usepackage{glossaries}

\usepackage{pgfplots} 

\pgfplotsset{compat = newest}

\newcommand{\figref}[1]%
{Figure \ref{#1}%
}

\newcommand{\tableref}[1]%
{Table \ref{#1}%
}

\newcommand{\algorithmref}[1]%
{Algorithm \ref{#1}%
}

\newcommand{\sectionref}[1]%
{Section \ref{#1}%
}

\newcommand{\lineref}[1]%
{Line \ref{#1}%
}

\algnewcommand{\LineComment}[1]{\State \(\triangleright\) #1}

\usepackage{newunicodechar}
\newunicodechar{−}{-}

\title{Idealize - A Notion of Idea Strength}

\author{Rui Portocarrero Sarmento \\
LIAAD-INESC TEC \\
PRODEI - Faculty of Engineering, University of Porto \\
mail@ruisarmento.com
}

\makenoidxglossaries

\newglossarystyle{quotes}{%
  \setglossarystyle{index}%
}

\begin{document}


\maketitle

\begin{abstract}
   Business Entrepreneurs frequently thrive on looking for ways to test business ideas, without giving too much information. Recent techniques in startup development promote the use of surveys to measure the potential client's interest. In this preliminary report, we describe the concept behind Idealize, a Shiny R application to measure the local trend strength of a potential idea. Additionally, the system might provide a relative distance to the capital city of the country. The tests were made for the United States of America, i.e., made available regarding native English language. This report shows some of the tests results with this system. 
\end{abstract}


\section{Introduction}

Business Idea strength is a function of several variables. One of these variables, the local strength, is a major variable when entrepreneurs are dealing with establishing a business. Some businesses vary the weight they consider this variable to have. For example, retail might give more weight to this than services. 

With the advent of internet search at the beginning of the XXI century, clients rely more and more on the Google search engine. This service provides locally oriented searches, and provide a list of available businesses that might fulfill the potential client needs.

In this report we test several assumptions, based on Google search engine and an index of local keyword strength, given an input summary for a business idea.

This document is organized in the following manner:
Section \ref{BACK} presents a small summary of previous work in this area. Then, in section \ref{TOOLS}, we present the tools we gathered to develop the testing system. Section \ref{DEV} presents some of the features of the developed system. Section \ref{RES} gives an idea of what results to expect from the system. Finally, in section \ref{DIS}, we discuss the main advantages and drawbacks of the solutions and give some suggestions for future work. Section \ref{CONC} concludes this document.

\section{Background}\label{BACK}

Some works have been done in the attempt to use some measures of search to improve prediction of outcome in several areas.

From \cite{BibEntry2019Mar}, we can read that some researchers stress that Google Trends data needs to be interpreted in the context in which a keyword is used, as well as the overall context of the research carried out. 

In the context of financial research, some authors have suggested that Google Trends data can be interpreted as a description of collective behavior, aggregate demand, stock market moves, investor expectations, information demand, attention, or market sentiment \citep{Curme11600,RePEc:chb:bcchwp:588,preis13,enlighten73287,RePEc:zbw:rwirep:155,RePEc:eee:jbfina:v:36:y:2012:i:6:p:1808-1821,doi:10.1111/j.1475-679X.2012.00443.x,RePEc:bla:ecorec:v:88:y:2012:i:s1:p:2-9}. 

Some of the quoted authors in \cite{BibEntry2019Mar} refer, as examples:

\begin{itemize}
    \item ``Internet search data may offer new possibilities to improve forecasts of collective behavior''
    \item ``Our findings /.../ suggests that Google data is a promising source of information for nowcasting components of aggregate demand in short-run models''
    \item ``By analyzing changes in Google query volumes for search terms related to finance, we find patterns that may be interpreted as 'early warning signs' of stock market moves.''
    \item ``We use a novelty Google search volume to proxy the market expectation hypothesis according to which firms with an abnormal upward change in Google searches are identified as firms with potential merger activity.''
    \item ``[W]e introduce a new indicator for private consumption based on search query time series provided by Google Trends."
    \item ``Demand is approximated in a novel manner from weekly internet search volume time series drawn from the recently released Google Trends database."
    \item ``We propose a new and direct measure of investor attention using search frequency in Google"
    \item ``We use daily internet search volume from millions of households to reveal market-level sentiment."
    \item ``How to use search engine data to forecast near-term values of economic indicators. Examples include automobile sales, unemployment claims, travel destination planning, and consumer confidence."
\end{itemize}

\section{Tools}\label{TOOLS}

In this section, we introduce some of the techniques we used in our tests. From software to concepts explored, everything is cited, and the methodology is described here. 

\subsection{TextRank}

\begin{figure}[H]
\centering
\includegraphics[scale=0.4]{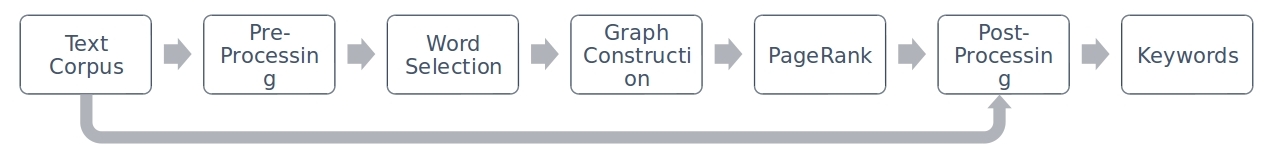}
  \caption{Original TextRank workflow}
  \label{fig:TRankDiag}

\end{figure}

In \cite{Mihalcea04TextRank}, the authors present the developed TextRank, as a solution to obtain keywords automatically from text. \figref{fig:TRankDiag} shows the workflow of the original TextRank algorithm.
The text corpus processing starts by the pre-processing of the text and the removal of stop words, numbers and punctuation. Then, the document goes through a process of annotation where remaining single words are categorized, for example, like nouns, verbs or adjectives among others. This method is called Part-of-Speech tagging (POS tagging). According to the authors, only a few of these annotated words are essential. The authors studied which group of words delivered the best results, and they concluded that the best automatic keyphrases were obtained with nouns and adjectives.
Then, with these filtered words, a graph-based approach is used. Each word is considered a graph node and the connections of words in this directed graph is determined by the order they appear in the text. The weight of these links is obtained by counting the number of times these pairs of words occur in the text corpus.
The next phase of the algorithm regards the selection of the words of high importance. This is done with the use of the PageRank algorithm by \cite{Pageetal98}. The words with high PageRank values are selected as potential keywords.
Finally, the keyphrases are obtained with a post-processing stage. This stage involves the use of a sliding window evolving through the initial text to assess the order of words that are contained in the keyphrases or keywords. This step takes into account punctuation and other structural features of the document to retrieve reasonable keyphrases.
TextRank has been widely praised as a consistent method to automatically retrieve keywords from the text. Inclusively, it has been used in prototypes of decision support systems as narrated by \cite{brazdil2015affinity}.

\subsection{Google Trends}

We used an R package (gtrendsR) for retrieving Google trends regarding input keywords. This makes a potentially good notion of the trend, in search hits, a particular keyword has. For more information, please see \cite{gtrendsR}.

Interest over time, a measure retrieved from Google Trends, is calculated as follows:

\begin{equation}\label{eq:intkey}
    {\displaystyle \operatorname {Interest} (Keyword)_{t}={\frac {\#of queries for  keyword_{t}}{Total Google Search Queries}}}
\end{equation}

Search interest is both indexed and normalized. This means the particular interest at time $t$, is divided by the maximum number of interest in the interval of search $x$. Equation \ref{eq:intkey} becomes equation \ref{eq:normintkey}, and this is how Google retrieves results of interest over time for a particular keyword.

\begin{equation}\label{eq:normintkey}
    {\displaystyle \operatorname {Normalized Interest} (Keyword)_{
    t \in {[t-x,t[}}={\frac {\#of queries for  keyword_{
    t\in {[t-x,t[}}}{{\max{\#of queries for keyword}_{[t-x,t[}}}*100}}
\end{equation}

For this reason, with regional analyses, we are retrieving a normalized indication of search interest within that particular country. An interesting index of 100 in Portugal and an index of 55 in Spain, would mean that the concentration of Portuguese searching for a particular keyword is higher than the concentration of Spanish searching for the same keyword. It can mean that Spanish are less interested in the keyword, or they may search for way more other concepts. The measure doesn't take into account the difference between countries in the size of the internet population and volume of queries per user.

\subsubsection{Normalized Keyword Weight}

Since the TextRank algorithm retrieves several keywords or keyphrases, we can achieve a weight we can use in the calculus of average trend; we will approach this in next subsections. Therefore, we use a Normalized Keyword Weight (NKW), from the TextRank algorithm:

\begin{equation}
    {\displaystyle \operatorname {NKW_{
    keyword_{k}}}={\frac{Weight_{keyword_{k}}}{{\sum_{k=1}^{nkeywords}{Weight_{keyword_{k}}}}}}}
\end{equation}\label{normintkey}

\subsubsection{Average Trend per Idea}

We achieve an average trend of our idea by adding weighted trend values per keyword, in the interval selected for search. The weight is retrieved from the TextRank algorithm, that gives us the weighted keywords or Keyphrases, given by TextRank weights. Thus, 

\begin{equation}
    {\displaystyle \operatorname {Average Trend Per Idea (Idea)_{
    t_{n}}}={\frac{\sum_{k=1}^{nkeywords}{NKW_{
    keyword_{k}}*NormalizedInterest(keyword_{k})_{t_{n}}}}{nkeywords}}}
\end{equation}\label{normintkey}

\section{Developed System}\label{DEV}

\begin{figure}[h]
\centering
\includegraphics[scale=0.50]{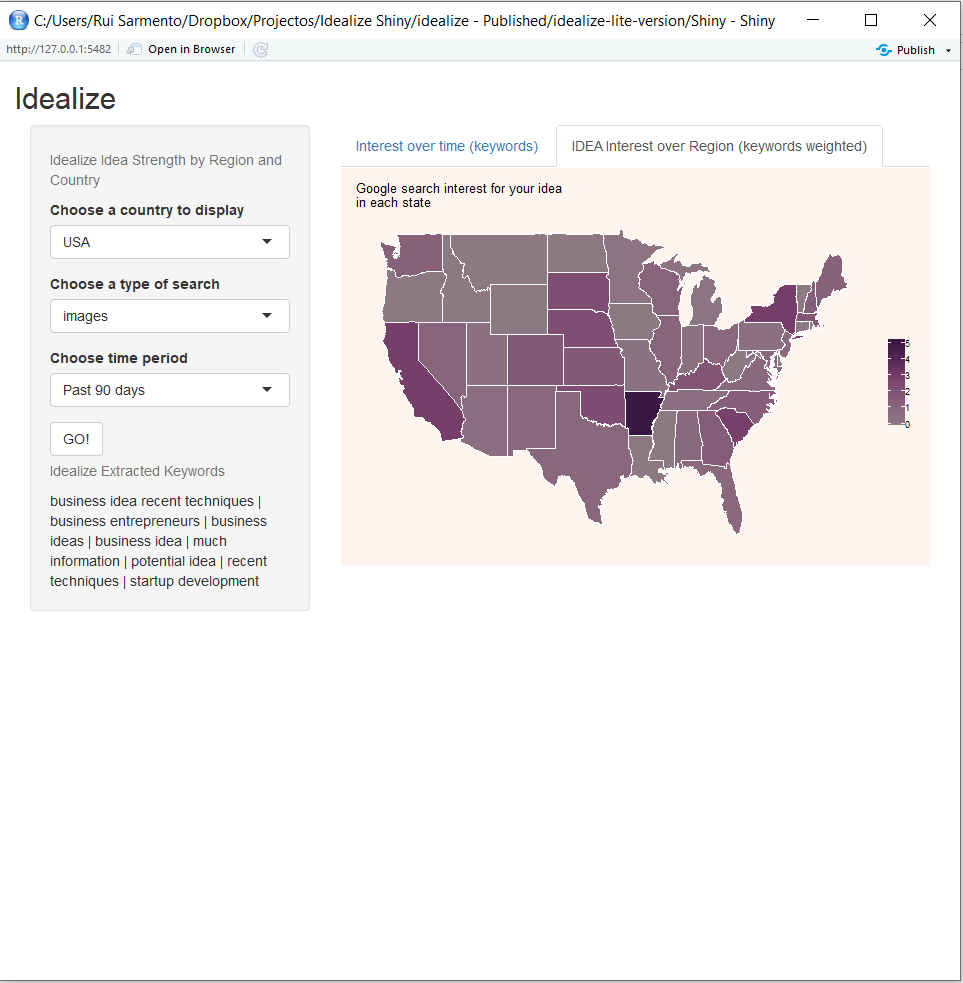}
  \caption{Idealize Application}
  \label{fig:Tudo}

\end{figure}

In figure \ref{fig:Tudo}, we can see the Idealize application \footnote{Available Code at \url{https://github.com/Sarmentor/idealize-lite-version}.}. There are two main regions in the application interface, one on the left of the user and another to the right of the screen.

\begin{figure}[h]
\centering
\includegraphics[scale=0.30]{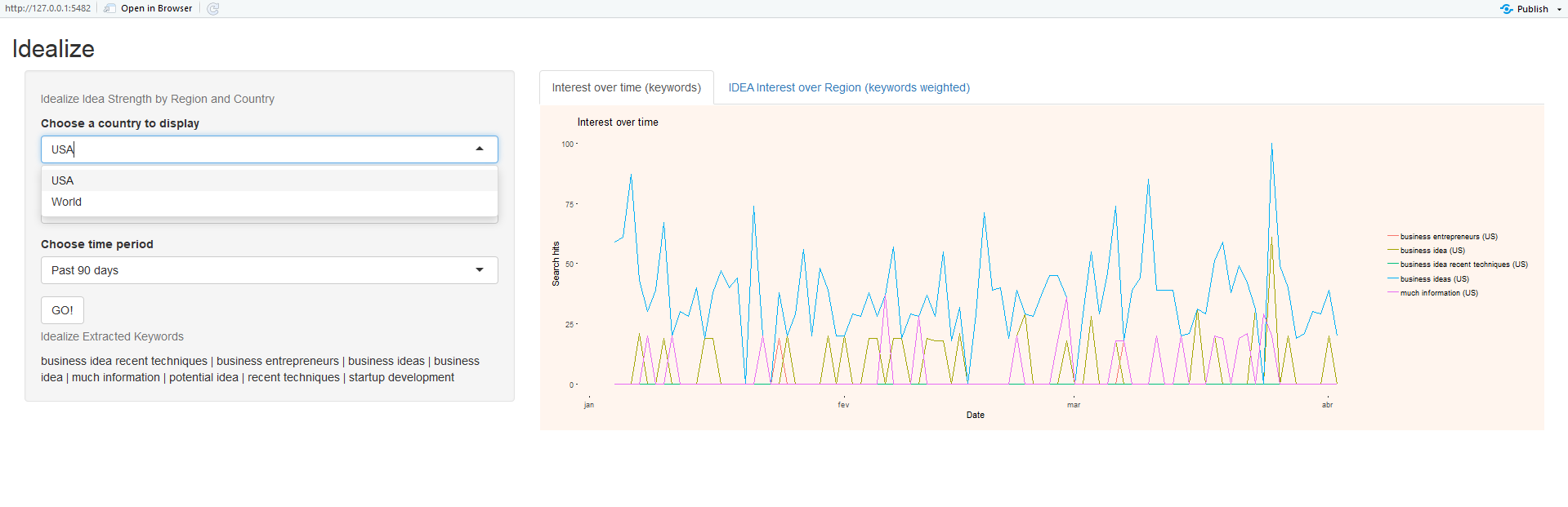}
  \caption{Idealize Application - Country Selection}
  \label{fig:Country}

\end{figure}

\begin{figure}[h]
\centering
\includegraphics[scale=0.30]{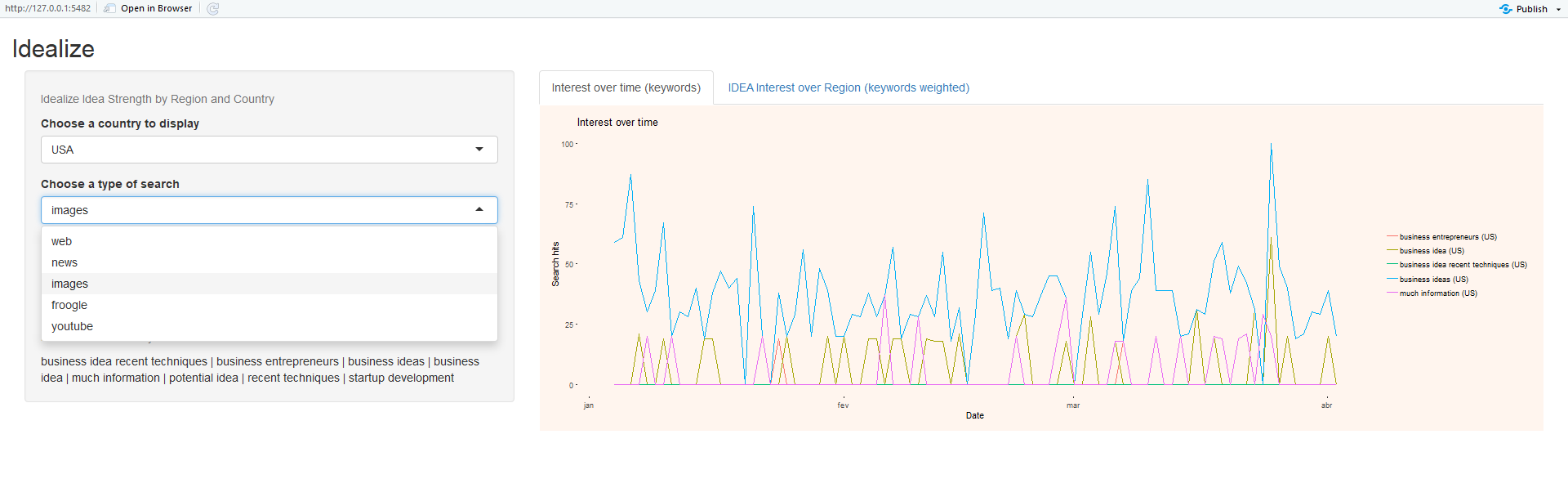}
  \caption{Idealize Application - Search Context Selection}
  \label{fig:Context}

\end{figure}

\begin{figure}[h]
\centering
\includegraphics[scale=0.30]{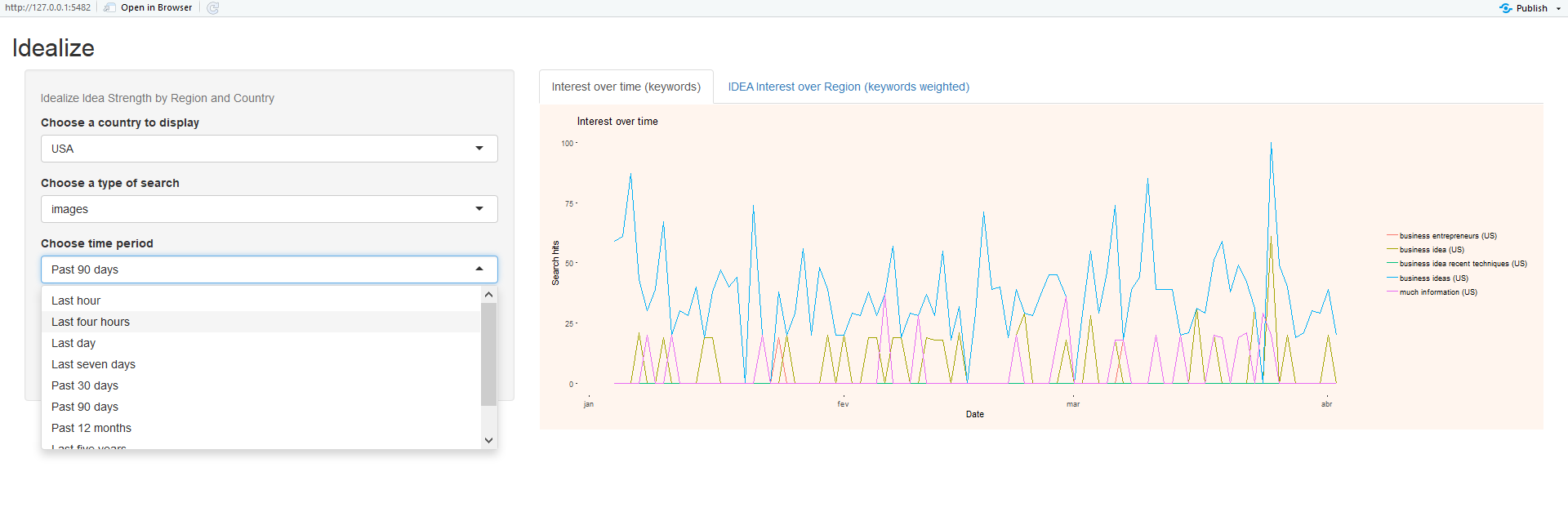}
  \caption{Idealize Application - Timeframe Selection}
  \label{fig:Timeframe}

\end{figure}

On the left region  (figures \ref{fig:Country},\ref{fig:Context} and \ref{fig:Timeframe}), the user can select a country or the world, regarding region of study. The user can also select the context of a search for the keywords extracted from the text, and finally, the user can select the timeframe for the searches.

Regarding Context Selection, the user can select one of the following choices:

\begin{itemize}
    \item ``web"
    \item ``news"
    \item ``images"
    \item ``froogle"
    \item ``youtube"
\end{itemize}

Regarding Timeframe Selection, the user can select one of the following choices:

\begin{itemize}
    \item ``Last hour"
    \item ``Last four hours"
    \item ``Last day"
    \item ``Last seven days"
    \item ``Past 30 days"
    \item ``Past 90 days"
    \item ``Past 12 months"
    \item ``Last five years"
    \item ``Since the beginning of Google Trends (2004)"
\end{itemize}

\section{Results}\label{RES}

\subsection{Idea Examples}

As an example, the input text was changed with two tentative concepts/ideas. 
We started by inputting this report's abstract. Then we inputted another conceptual idea, this time regarding automobile services. From the keywords in the texts, we extracted plots of keyword trends, and also the plots for the idea strength per region of study.

\subsubsection{Input Text I}

``Business Entrepreneurs frequently thrive on looking for ways to test business ideas, without giving too much information about their business idea. Recent techniques in startup development promote the use of surveys to measure the potential client's interest in a business idea. In this preliminary report, we describe the concept behind Idealize, a Shiny R application to measure the local trend strength of a potential idea for a business. Additionally, the system might provide a relative distance to the capital city of the country. The tests were made for the United States of America, i.e., made available regarding native English language. This report shows some of the tests results with this system.''

\paragraph{Trends and maps for the USA \newline\newline}

In Figures \ref{fig:Ab5WebTrend} and \ref{fig:Ab5WebIdea} (see APPENDIX), we show the results for this text, regarding a 5 year timeframe and for web searches.

Figure \ref{fig:Ab5WebTrend} shows the trend chart of several extracted keywords; it is relatively clear that some keywords have different trends, some more seasonal than others. It is also visible we show only five keywords. This is due to a limitation in the gtrendsR package that allows a limit of 5 keywords searches at a time.

Figure \ref{fig:Ab5WebIdea} shows the idea strength map. It is shown that some states give more importance to such an idea than others. As darker the color in the selected color gradient, the more importance is given to the user idea.

In Figures \ref{fig:Ab1YouTrend} and \ref{fig:Ab1YouIdea} (see APPENDIX), we show the results for this same text, this time regarding a 1-year timeframe and Youtube searches.

Figure \ref{fig:Ab1YouTrend} shows the trend chart of the several extracted keywords. As expected, keywords have different trends, again, some more seasonal than others.

Figure \ref{fig:Ab1YouIdea} shows the idea strength map. It is shown the importance that states give to such an idea.

\subsubsection{Input Text II}

``Our company provides the best service in auto maintenance and tunning. We have several stores around the country. We serve our clients all auto brands, either regarding car parts or maintenance.

In the tunning department, we provide several tunning brands and also provide tunning services as per the client needs.''

\paragraph{Trends and maps for the USA \newline\newline}

In Figures \ref{fig:Au1WebTrend} and \ref{fig:Au1WebIdea} (see APPENDIX), we show the results for this text, regarding a 1-year timeframe and web searches.

Figure \ref{fig:Au1WebTrend} shows the trend chart of the several extracted keywords; it is relatively clear that some keywords have different trends, comparatively with the first idea results.

Figure \ref{fig:Au1WebIdea} shows the second idea strength map. Comparatively, with the first idea, we can see that with the idea concept change, several changes were produced in the distribution, per state, regarding idea strength. This is the expected result, with the use of this system.

\section{Discussion}\label{DIS}

This report brings some light to some possible methods to approach idea development from its inception. Although several other approaches are possible, this one involves very low effort. A logical synopsis or summary of an idea and a location is all that is needed. Nonetheless, as reported, this system has several concepts that are far from perfect; for example, there are some base measures that are simple approaches and not exact. Moreover, some values change subjectively, with variations of text input and English language grammar. This is true, although the base conceptual idea to be tested might be the same, but explained with different words.

Additionally, regarding the evaluation of the model, we did not have the time or resources available to retrieve ground truth data to test our assumptions. This would be possible, for example, by accessing startups or SME's temporal data of results. This would be important to fine tune some tasks in the several phases of the workflow in the proposed system.

\section{Conclusions}\label{CONC}

This academic report, about the experiment with idea strength measurement, brought light to some methods for achieving a notion about an idea of business. This proposed low-cost method is interesting but exposes many issues or doubts about its value. There are many variables that might be added in the future to such a system. No system is perfect, and the complexity of such a prediction as idea strength has many factors involved. The presented solution does not entitle itself as the final solution, as discussed in this report, many other factors might have to be used to achieve a good prediction of idea strength.

\section*{Acknowledgments}
This work was fully financed by the Faculty of Engineering of the Porto University. Rui Portocarrero Sarmento also gratefully acknowledges funding from FCT (Portuguese Foundation for Science and Technology) through a Ph.D. grant (SFRH/BD/119108/2016). The authors want to thank also to the reviewers for the constructive reviews provided in the development of this publication.

\bibliographystyle{apalike}
\bibliography{Report}

\cleardoublepage
    \appendix
\chapter{APPENDIX}
\section{RESULTS FIGURES}

\subsection{IDEA I}

\begin{figure}[H]
\centering
\includegraphics[scale=0.35]{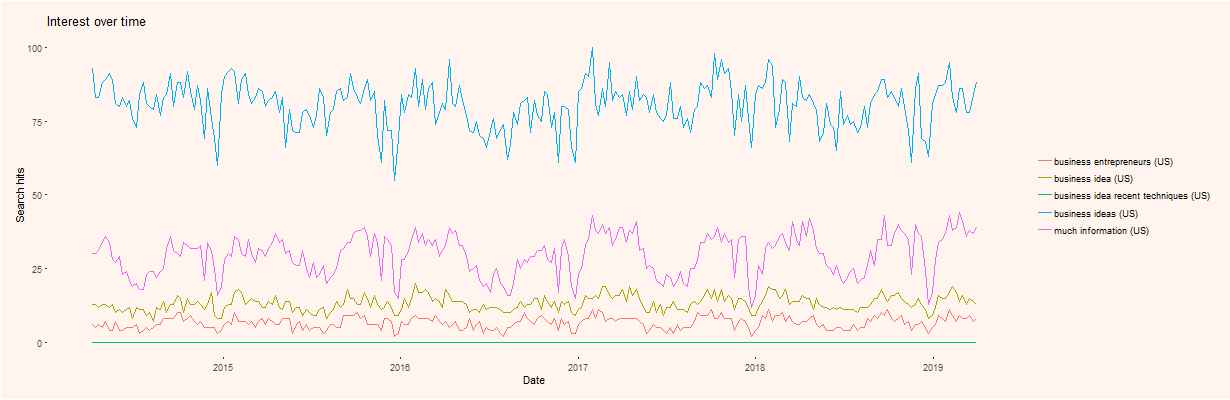}
  \caption{Abstract Concept Idea - Keyword Trends - for last 5 years (web searches)}
  \label{fig:Ab5WebTrend}

\end{figure}

\begin{figure}[H]
\centering
\includegraphics[scale=0.55]{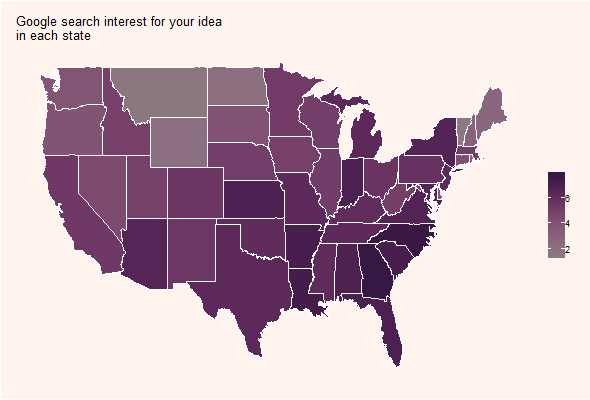}
  \caption{Abstract Concept Idea - Idea Strength per State Region - for last 5 years (web searches)}
  \label{fig:Ab5WebIdea}

\end{figure}

\begin{figure}[H]
\centering
\includegraphics[scale=0.35]{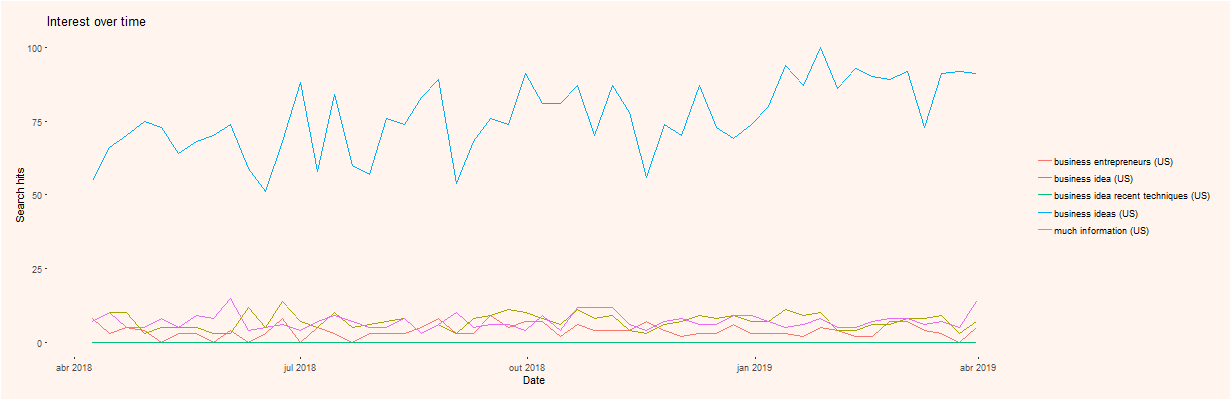}
  \caption{Abstract Concept Idea - Keyword Trends - for last 1 year (Youtube searches)}
  \label{fig:Ab1YouTrend}

\end{figure}

\begin{figure}[H]
\centering
\includegraphics[scale=0.55]{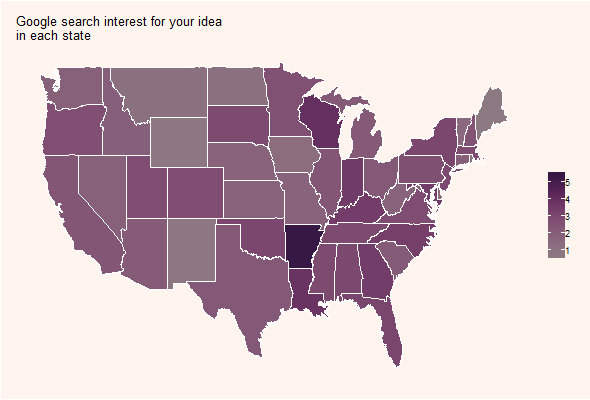}
  \caption{Abstract Concept Idea - Idea Strength per State Region - for last 1 year (Youtube searches)}
  \label{fig:Ab1YouIdea}

\end{figure}

\subsection{IDEA II}

\begin{figure}[H]
\centering
\includegraphics[scale=0.35]{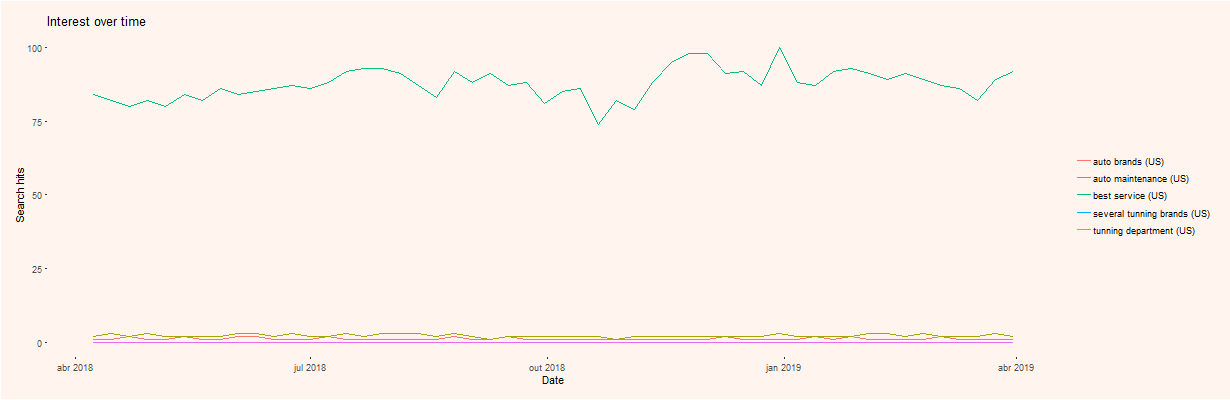}
  \caption{Auto Concept Idea - Keyword Trends - for last 1 year (web searches)}
  \label{fig:Au1WebTrend}

\end{figure}

\begin{figure}[H]
\centering
\includegraphics[scale=0.55]{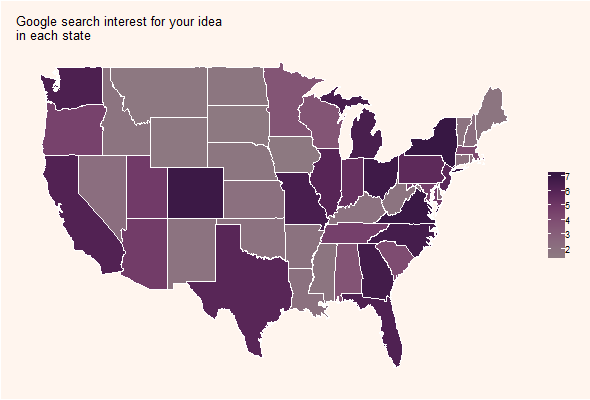}
  \caption{Auto Concept Idea - Idea Strength per State Region - for last 1 year (web searches)}
  \label{fig:Au1WebIdea}

\end{figure}
\cleardoublepage

\end{document}